\documentstyle [epsfig,wrapfig,11pt]{article}
\begin{document}
\hfill  ORNL-CCIP-93-08 / RAL-93-062
\vspace{1.0cm}
{
\begin{center}
{\Large\bf Hadronic Molecules and Scattering Amplitudes\\
from the Nonrelativistic Quark Model}
\footnote[1]{Presented at HADRON93, Como, Italy (21-25 June 1993).}\\
\vspace{1cm}
T.Barnes\\
Physics Division and Center for Computationally Intensive Physics\\
Oak Ridge National Laboratory, Oak Ridge, TN 37831-6373\\

\date{}
\end{center}
}

\begin{abstract}
This report summarizes recent calculations of
low-energy hadron-hadron scattering
amplitudes in the nonrelativistic quark potential model, which assume that the
scattering mechanism is a single interaction
(usually OGE) followed by constituent interchange.
We refer to the scattering diagrams as ``quark Born diagrams". For the cases
chosen to isolate this mechanism, I=2 $\pi\pi$, I=3/2 K$\pi$, KN and NN,
the results are usually in good agreement with experimental S-wave scattering
amplitudes given standard potential-model parameters. These calculations also
lead to predictions of vector-vector bound states, one of which may be the
$\theta(1710)$. This assignment can be tested by searches for
K$\bar{\rm K}\pi\pi$
and $\phi\pi^0\gamma$ decay modes of the $\theta(1710)$.

\end{abstract}

\section{Introduction}
\thispagestyle{empty}

The determination of the strong forces between hadrons in terms of the
underlying quark and gluon degrees of freedom has been a long-standing goal of
research in QCD. Many theoretical studies at the quark-and-gluon level have
considered the important nucleon-nucleon interaction \cite{hyperf}, and have
been successful in extracting the short-range repulsive core, which they find
to be dominated by the OGE spin-spin interaction.
Interactions of pseudoscalar mesons have been studied variationally
in the nonrelativistic quark potential model \cite{WI}; these calculations
found qualitatively correct results for S-wave phase shifts, and predicted
deuteronlike kaon-antikaon ``molecule" bound states, which have been identified
with the $f_0(975)$ and $a_0(980)$ just below K$\bar{\rm K}$
threshold. Although the
results of these studies are very encouraging, the nonperturbative
techniques used (resonating group methods and many-parameter variational
calculations) require considerable theoretical effort to determine
hadron scattering amplitudes.

Our collaboration has investigated the possibility that certain low-energy
hadron-hadron scattering reactions may be dominated by simple perturbative
processes. We consider channels in which s-channel resonances are not expected,
and (except for the baryon-baryon case) will discuss reactions involving an
external pseudoscalar meson, so one-pion-exchange is forbidden. Through these
constraints we hope to isolate and study the mechanism of hadron scattering
which does not involve valence annihilation and pair production.

\section{Method}

We assume that nonresonant hadron scattering is dominated by the lowest-order
perturbative QCD process, specifically one-gluon-exchange followed by
constituent (quark) interchange. Nonperturbative QCD is assumed to contribute
to scattering only through the formation of the asymptotic hadron wavefunctions
(or in the most detailed study as a single scattering interaction through a
linear confining potential \cite{Swan}). The constituent-interchange scattering
mechanism is of course not new \cite{CI}. Our contribution to this picture is
to couple constituent interchange to OGE in the nonrelativistic quark potential
model, and to evaluate the color, spin and spatial matrix elements within this
OGE+CI formalism.
We use a conventional nonrelativistic quark-model Hamiltonian of the form
\begin{equation}
H = \sum_{i=1}^4 \left( {p_i^2 \over 2 m_i} + m_i \right) + \sum_{i<j}
\left[ V_{conf}(r_{ij}) + V_{hyp}(r_{ij})\; \vec S_i \cdot \vec S_j
\; \right] (\lambda^a_i/2) \cdot (\lambda^a_j/2) \ ,
\end {equation}
\noindent
where
$V_{hyp} = -(8 \pi \alpha_s / 3 m_i m_j)\; \delta(\vec r_{ij})$
is the contact color-hyperfine interaction and
$V_{conf}$ is the spin-independent
confining potential. We then calculate the Born-order scattering amplitude
and corresponding phase shifts, using relativistic phase space and
kinematics for the external hadrons.
In most cases discussed here the hyperfine
term dominates \cite{Swan}, and we show explicit results for this interaction
and use simple Gaussian hadron wavefunctions for illustration unless stated
otherwise.

In a two-meson scattering process without identical quarks the OGE + CI
scattering mechanism leads to four scattering diagrams, since there are $2
!\cdot 2 ! $ permutations of OGE interactions between the
incoming lines. One such diagram is shown below, and the rules for
generating and evaluating these diagrams are given in reference \cite{BS}.

\setlength{\unitlength}{1.5pt}
\begin{picture}(180,70)(-50,-5)
\put(30,30) {\makebox(0,0)[1]{{\it capture}$_1 \ \ $  = } }
\put(75,50) {\makebox(0,0)[1]{A} }
\put(75,10) {\makebox(0,0)[1]{B} }
\put(160,10) {\makebox(0,0)[1]{D.} }
\put(160,50) {\makebox(0,0)[1]{C} }
\put(76,0){
\begin{picture}(75,60)(-2,0)
\multiput(5,15)(0,40){2}{\multiput(0,0)(55,0){2}{\vector(1,0){5}}}
\multiput(15,5)(0,40){2}{\multiput(0,0)(55,0){2}{\vector(-1,0){5}}}
\multiput(5,5)(60,10){2}{\multiput(0,0)(0,40){2}{\line(1,0){5}}}
\multiput(10,15)(0,40){2}{\line(1,0){50}}
\multiput(15,5)(0,40){2}{\multiput(0,0)(35,0){2}{\line(1,0){15}}}
\put(30,5){\line(1,2){20}}
\put(30,45){\line(1,-2){20}}
\put(20,5){\dashbox{2}(0,50){}}
\multiput(20,5)(0,50){2}{\circle*{2}}
\end{picture}
}
\end{picture}

\section{Results for $\pi\pi$, K$\pi$, VV, KN and NN}

\noindent
{\it a) I=2 $\pi\pi$}
\vskip 0.2cm

\begin{wrapfigure}{r}{3.0in}
\epsfig{file=h1.epsm,width=3.0in}
\caption{I=2 $\pi\pi$ S-wave phase shift.}
\label{fig 1}
\end{wrapfigure}
The I=2 $\pi\pi$ scattering amplitude due to the contact spin-spin interaction
with Gaussian wavefunctions may be evaluated in closed form
\cite{BS}. The S-wave is
dominant at energies studied experimentally, and is given by
(2) below.
The two free parameters
are reasonably well established in quark model
phenomenology, $\alpha_s/m_q^2 \approx (0.6/0.33^2)$ GeV$^{-2}$ and $\beta_\pi
\approx 0.3 $ GeV. We fix $\alpha_s/m_q^2$ at this standard value and fit the
less well-determined Gaussian width parameter $\beta_\pi$ to the data of
Hoogland {\it et al.} \cite{Hoog}. This is shown in figure 1, and the
fitted value is $\beta_\pi=0.337$ GeV, close to expectations. There is a clear
need for an experimental determination of the phase shift above $M_{\pi\pi}=
1.5$ GeV, which we predict to be near an extremum.

The S-wave phase shift is
\begin{equation}
\sin \delta_2^{(0)}
= - \Bigg\{
{4\alpha_s \beta_\pi^2 \over 9 m_q^2 }
\;
\Bigg( 1 - e^{-(s-4M_\pi^2)/8\beta_\pi^2 }
+{(s-4M_\pi^2)\over \sqrt{27}\beta_\pi^2  }\,
e^{-(s-4M_\pi^2)/12\beta_\pi^2  }  \,
\Bigg) /
\sqrt{ 1-4M_\pi^2/s}
\Bigg\}
\ ,
\end{equation}
which leads to an
I=2 $\pi\pi$ scattering length of
\begin{equation}
a_2^{(\pi\pi)} = -{2\over 9} \bigg(1 + {8\over \sqrt{27}} \bigg)
{\alpha_s M_\pi \over m_q^2} \ .
\end{equation}
This is numerically
$a_2^{(\pi\pi)} /  M_\pi^{-1}   =  -0.059$,
in good agreement with Weinstein's (unscaled) variational result
\cite{JW}, and with the PCAC result of Weinberg \cite{Wei} and the
recent parametrization of Morgan and Pennington \cite{MP}; the latter
two references both find
$a_2^{(\pi\pi)} / M_\pi^{-1} = -0.06 $.
\vskip 0.2cm
\noindent
{\it b) I=3/2 } K$\pi$
\vskip 0.2cm

Another annihilation-free channel is I=3/2 K$\pi$
\cite{BSW}. Here we have four free parameters,
$\alpha_s/m_q^2$, $m_q/m_s$, $\beta_\pi$ and $\beta_K$. We leave
$\alpha_s/m_q^2$ and $\beta_\pi$ equal to their $\pi\pi$ values. The
phase shifts are found to be rather insensitive to the relative kaon/pion
length scale, so we set $\beta_K=\beta_\pi$; this leaves only
$m_q/m_s$ undetermined, and typical constituent quark parameters suggest
$m_q/m_s\approx 0.33$ GeV/0.55 GeV = 0.6.

\begin{wrapfigure}{r}{3.5in}
\epsfig{file=h2.epsm,width=3.5in}
\caption{I=3/2 K$\pi$ S-wave and P-wave phase shifts.}
\label{fig 2}
\end{wrapfigure}

When we use the S-wave phase shift
data of
Estabrooks {\it et al.} \cite{Esta} to determine $m_s$
we find that the optimum ratio
is $m_q/m_s=0.677$, similar to expectations.
Imposing $m_q\neq m_s$ has an interesting effect; it induces a P-wave
amplitude.
The theoretical S-wave and P-wave phase shifts are shown in figure 2. Note that
the previous prediction of an extremum in the I=2 $\pi\pi$ S-wave phase shift
near 1.5 GeV (in figure 1) also holds for its SU(3)-partner I=3/2 K$\pi$ S-wave
phase shift (figure 2). I=3/2 K$\pi$ experimental data are available
at higher invariant mass than I=2 $\pi\pi$,
and appear to support this prediction. Our result for
the scattering length,

\begin{equation}
a_{3/2}^{(K\pi)} =   - { 2\alpha_s \over 9 m_q^2 (M_\pi^{-1} + M_K^{-1}) }
\left[ 1 +
  \left( {4 \beta_\pi^2 \over  2\beta_\pi^2 + \beta_K^2 }\right)^{3/2}
+  {m_q\over m_s} \Bigg\{ \left( {4 \beta_K^2  \over 2\beta_K^2  +\beta_\pi^2 }
\right)^{3/2} +
\left( {2 \beta_K\beta_\pi \over \beta_K^2  +\beta_\pi^2 }\right)^3
\Bigg\} \right]
\end{equation}
corresponds to about $-0.077/M_\pi$
with our preferred parameter set; this is
consistent with the PCAC
result of $\approx -0.07/M_\pi$ \cite{Wein} and with the (rather
wide) range of reported experimental values. Finally, the predicted and
observed P-waves are qualitatively consistent.

\vskip 0.2cm
\noindent
{\it c) Other meson-meson channels; vector-vector molecules}
\vskip 0.2cm

I=1 KK scattering may be treated using essentially identical techniques
\cite{BS}. We find a moderately strong repulsive interaction, which
agrees with the variational results of Weinstein \cite{JW}.
Unfortunately there are no experimental results for this reaction.

Although these analytical techniques
have not yet been applied in detail to other
meson-meson channels, Swanson \cite{Swan} has carried out similar numerical
studies of the phenomenologically interesting vector-vector and
vector-pseudoscalar channels with $u,d$ and $s$ quarks. The interesting
question in these channels is whether unusual states such as the $\theta(1710)$
and the $f_1(1420)$ might actually be weakly-bound molecular vector-vector
bound states or K$^*$K threshold enhancements, respectively.

Bound states form most easily in strongly-coupled systems that are nearly
degenerate, since there are small energy denominators, and one linear
combination of states in a $2\times 2$ Hamiltonian with an off-diagonal
interaction always experiences an attractive interaction. The $qs\bar q \bar s$
vector-vector system is such a case; here Swanson finds an especially large
K$^*\bar{\rm K}^* \leftrightarrow \omega\phi$ matrix element, which in a
multichannel Schr\"odinger formalism described by Dooley, Swanson and Barnes
\cite{DSB} leads to scalar and tensor bound states. These vector-vector
molecules might be identified with the $\theta(1710)$; similar suggestions for
this and other unusual resonances have appeared in the literature
\cite{Kala,Torn,EK}.
The couplings and branching fractions of
the linear combination
$|\theta\rangle = (|{\rm K}^*\bar {\rm K}^*\rangle +
|\omega\phi\rangle )/\sqrt{2}$
which is formed by a large
off-diagonal interaction explain
many of the unusual
properties of the $\theta$.
New predictions of this model
\cite{DSB} include a large branching fraction to
K$\bar {\rm K}\pi\pi$ ($B.F.\approx 35\% $),
an unusual $\phi\pi^0\gamma$ mode ($B.F.\approx 0.3\% $),
$B.F. {\rm K}\bar {\rm K} / B.F. \eta\eta \approx 2$,
a weak $\pi\pi$ mode ($B.F.\sim 5-10\% $),
the $\psi$-hadronic ratio
$(\psi\to\phi\theta) / (\psi\to\omega\theta) \approx 2/\lambda^2$
(just as for
a ${\rm K}\bar {\rm K}$ molecule), and a small
$\gamma\gamma$ width \cite{thetagams}.

\vskip 0.2cm
\noindent
{\it d) }KN
\vskip 0.2cm

One may also study meson-baryon and baryon-baryon scattering using quark Born
diagrams. We have considered KN scattering in detail \cite{BSKN}, since there
are no valence annihilation contributions and the low partial waves have been
studied experimentally. In KN scattering there are I=0 and I=1 channels, and
both are repulsive, with I=1 having the strongest interaction.

\begin{wrapfigure}{r}{3.0in}
\epsfig{file=h3.epsm,width=3.0in}
\caption{KN S-wave phase shifts; I=0 (open points) and I=1 (solid points).}
\label{fig 3}
\end{wrapfigure}
These features of KN interactions are correctly predicted
by the quark Born formalism. The
well-determined I=1 KN scattering length of  $\approx -0.33$ fm is predicted to
be about $-0.35$ fm given standard quark model parameters (dashed).
The I=0 scattering
length is predicted to be about $-0.12$ fm, but is unfortunately not yet well
determined experimentally. At higher energies we find that the Born-order
S-wave phase shifts are too ``soft"; the predicted phase shifts are largest at
$P_{cm}\approx 0.5$ GeV, whereas the observed ones appear to grow
monotonically to the maximum experimental momentum of $P_{cm}\approx 0.7$ GeV.
This may be an artifact of our ``soft" Gaussian baryon wavefunctions, since a
good fit is possible given a moderately reduced nucleon length scale (solid).
Higher KN partial
waves are also interesting in that they show clear evidence of spin-orbit
interactions, which we have not yet incorporated in our formalism.

\vskip 0.2cm
\noindent
{\it e) }NN
\vskip 0.2cm

Finally, the short-range repulsive core in the NN interaction provides
an important test of any description of
hadron-hadron scattering. We assume that
the repulsive core is dominantly due to the OGE spin-spin hyperfine
interaction, which was a conclusion of resonating group
and variational studies
\cite{hyperf}. We do not expect to reproduce the longer-ranged attraction,
which is variously attributed to spatial wavefunction distortion or meson
exchange; neither effect is present in our formalism at Born order.

When modelled as a potential the short-distance core is typically found to have
a maximum of $\approx 1$ GeV and a range of about 1/2 fm for both NN isospin
states. In our formalism this is a straightforward calculation \cite{BSNN}, and
the two parameters $\alpha_s/m_q^2$ and $\alpha_N$ (baryon
Gaussian width parameter)
are already determined. The
low-energy NN potentials we find are shown in figure 4.

\begin{wrapfigure}{r}{3.0in}
\epsfig{file=h4.epsm,width=3.0in}
\caption{I=0 and I=1 low-energy NN potentials.}
\label{fig 4}
\end{wrapfigure}
These
core potentials are very similar to the results of resonating
group calculations. Thus, our rather surprising conclusion is that the
cores are dominantly Born-order one-gluon-exchange effects. The
Born-order S-wave NN phase shifts in contrast are quite inaccurate,
since the potentials are so large.
We have used numerical integration of the Schr\"odinger equation to
determine phase shifts which result from the Born-order NN
core potentials, and the resulting
phase shifts closely resemble those of Oka
and Yazaki \cite{hyperf}.

In a search for possible molecule states we investigated other nonstrange
baryon-baryon channels. We found that the I=0, S=1 $\Delta\Delta$ channel has
an attractive core and might form a bound state. However, the assumptions of
hyperfine dominance and the single-channel approximation
have not been tested for this system and may modify our conclusion about
bound states, since a fall-apart coupling from
I=0, S=1 $\Delta\Delta$ to NN is present.

\section{Summary and Conclusions}

We have found that a simple perturbative mechanism, one-gluon-exchange followed
by constituent interchange, leads to an accurate description of low energy
hadron-hadron scattering in several annihilation-free channels, given standard
quark-model parameters. In future work it will be interesting to generalize
these techniques through
1) the inclusion of other terms such as spin-orbit (needed
for KN higher partial waves),
2) the use of more realistic baryon
wavefunctions (for KN at higher energies), and
3) the incorporation of $q\bar q$
annihilation, which is known to be a very important effect when allowed,
as in K$\bar {\rm K}$
and I=0 $\pi\pi$ scattering. With these improvements we expect
that it will be possible to predict phase shifts and bound states accurately in
other meson-meson and meson-baryon channels, which will be accessible to
experiments at hadron facilities such as CEBAF, DAPHNE, KAON, KEK and LEAR. Our
comparisons with experiment suggest that better determinations of some
scattering amplitudes, such as I=2 $\pi\pi$ (above 1.5 GeV), I=3/2 K$\pi$ and
especially I=0 KN, would be very useful contributions to the study of soft
hadron scattering.

\section{Acknowledgements}

It is a pleasure to acknowledge the efforts of the organisers of this meeting,
in particular the co-chairmen T.Bressani and G.Preparata. The advice and
essential contributions of my collaborators S.Capstick,
K.Dooley, G.Grondin, N.Isgur, M.D.Kovarik, E.S.Swanson and
J.Weinstein
are also gratefully acknowledged. This research was sponsored in
part by the United States Department of Energy under contract
DE-AC05-840R21400, managed by
Martin Marietta Energy Systems, Inc, and by the United Kingdom Science Research
Council through a Visiting Scientist grant at Rutherford Appleton Laboratory.

\end{document}